\documentstyle[prl,aps,preprint]{revtex}
\begin{document}
\setcounter{page}{0}
\def\footnoterule{\kern-3pt \hrule width\hsize \kern3pt}
\tighten

\title{Mass Spectrum for Black Holes in\\
  Generic 2-D Dilaton Gravity\footnote{Talk given by G. Kunstatter
at the 2nd Sakharov Conference, Moscow, 1996.}}

\author{A. Barvinsky}

\address{Theory Department,Lebedev Physics Institue \\
Leninsky Prospect 53, Moscow 117924, Russia \\
 {Email address: {\tt barvin@rq2.fian.msk.su}}}

\author{G. Kunstatter}

\address{Physics Department, University of Winnipeg \\
Winnipeg, Man., Canada R3B 2E9\\
{Email address: {\tt gabor@theory.uwinnipeg.ca}}}

\maketitle

\begin{abstract}

Two arguments  for the quantization of entropy
for black holes in generic 2-D dilaton gravity are summarized.
The first argument
is based on  reduced quantization of the only physical observables in the
theory, namely the black hole mass and its conjugate momentum, the Killing
time separation. The second one uses the exact physical mass eigenstates
for Euclidean black holes found via Dirac quantization. Both methods give the
same spectrum: the black hole entropy must
be quantized $S= 2\pi n/G$.
\end{abstract}

\clearpage

\par
Progress in the theory of quantum black holes and their thermodynamics
essentially depends on the resolution of the problem of Bekenstein-Hawking
entropy \cite{entropy}: the explanation of the statistical mechanical origin of
this geometric quantity. While the first law of thermodynamics for
black holes seems to be a very general and model independent concept
\cite{Wald}, the statistical mechanical confirmation of the corresponding
entropy exists only for a limited set of examples, such as extremal,
string-inspired supersymmetric black holes\cite{Tseytlin}, and black holes in
2+1
dimensions\cite{carlip}.
 The universality of black hole thermodynamics \cite{Wald},
however, suggests that the explanation of the entropy and energy spectra of
black holes might  not be found in  terms of specific models.
Instead one should perhaps search for quantum mechanical mechanisms of their
origin in the simplest (vacuum or eternal black hole) cases. One such
mechanism has recently been proposed on rather general and heuristic
grounds by Bekenstein and Mukhanov \cite{bekenstein} who argued that
the area $A(M)$ of a black hole with mass $M$
should be quantized according to the rule:
\begin{equation}
A(M) = \alpha n
\end{equation}
where the coefficient $\alpha$ was determined by requiring the degeneracy
of states:
\begin{equation}
g(M) = \exp S(M) = \exp (A(M) /4)
\end{equation}
be an integer. A spectrum of this form has also been obtained in a variety
of other contexts\cite{mazur,berezin,louko1}. In this talk we will summarize
recent arguments\cite{BK}
that suggest a similar quantization condition on
the entropy  $S(M)$ of  black holes with mass $M$ in generic
2-D dilaton gravity, namely:
\begin{equation}
S(M) = 2\pi n/G,
\end{equation}
but based on different quantum mechanical grounds. Since generic dilaton
gravity contains as a special case
spherically symmetric gravity, this quantization condition should, if valid,
also apply to Euclidean black holes in Einstein gravity.
\par
The classical action for generic dilaton gravity in 2 spacetime
dimensions is\cite{banks}
\begin{equation}
I = {1\over 2 G} \int dt dx \sqrt{-g} \left(\eta R(g)+
   {V(\eta)\over l^2}\right)
\label{eq: action1}
\end{equation}
where $G$ is a (dimensionless) gravitational constant, and $l$ is a fundamental
constant of dimension length.
 The most general solution to the field equations in the generic theory up
to spacetime diffeomorphisms can be written\cite{domingo2,DGK}:
\begin{eqnarray}
ds^2&=&-(j(\eta) - 2GlM)dt^2 + {1\over (j(\eta) - 2GlM)} dx^2
\nonumber\\
\eta &=& x/l
\label{eq: general solution}
\end{eqnarray}
where $j(\eta) = \int^\eta_0 d\tilde{\eta} V(\tilde{\eta})$. These solutions
all have a Killing vector, whose
norm can be written in coordinate invariant form as
\begin{equation}
|k|^2 = -l^2 |\nabla \eta |^2 = ( 2GlM - j(\eta))
\label{eq: killing vector}
\end{equation}
As long as $j(\eta)$ rises monotonically from zero at $\eta=0$, the above
solutions describe Schwarzschild-like black holes with ADM mass $M$. Note
that in the above parametrization the
metric is not asymptotically flat, nor is the Killing vector normalized to
one at spatial infinity. Without changing the essential features of the
following arguments, one can  define a new physical, asymptotically
flat metric by the conformal reparametrization $\tilde{g}_{\mu\nu} =
g_{\mu\nu}/j(\eta)$.
A complete discussion of the necessary conditions for the existence of black
hole solutions in the generic theory, and the corresponding thermodynamics,
can be found in \cite{DGK}. The crucial observation for the present purposes is
that event horizons are surfaces $\eta=\eta_-$= constant for which $|k|^2=0$.
The entropy of the corresponding black hole is proportional to
the value
of the dilaton at the horizon, namely
\begin{equation}
S= {2\pi\over G}\eta_-
\label{eq: entropy}
\end{equation}
\par
The Hamiltonian formulation for the geometrical theory starts with a
decomposition of the metric:
\begin{equation}
ds^2 = e^{2\rho}\left(-u^2dt^2 +(dx-vdt)^2\right)
\label{eq: metric decomp}
\end{equation}
The boundary conditions relevant to black hole thermodynamics require placing
the black hole in a box of fixed radius (the value of the dilaton at the
boundary is fixed). Moreover we restrict the slices on the interior of the
box to end at the bifurcation point of an eternal black hole (i.e. the point
at which the Killing vector vanishes), so that we can analytically continue
the solutions to Euclidean time without obstruction. As shown in \cite{LW}
for spherically symmetric gravity, and in \cite{BK} for the generic theory,
the resulting Hamiltonian is:
 \begin{equation}
 H_c= {\int^{\sigma_+}_{\sigma_-}dx}
   \left(-\tilde{u}{\cal M}'+\tilde{v}{\cal P}\right) + H_+-H_-
\label{eq: canonical hamiltonian1}
\end{equation}
where $\tilde{u}$ and $\tilde{v}$ are Lagrange multipliers and $\Pi_\rho$,
$\Pi_\eta$ are the momenta conjugate to $\rho$, $\eta$, respectively.
In the above
${\cal P} = \Pi_\rho' -\Pi_\rho \rho' - \Pi_\eta \eta'$ is the generator of
spatial diffeomorphisms, and the Hamiltonian constraint has been rewritten
as the spatial gradient of the mass observable
$\cal{M}$\cite{kuchar,domingo2,thiemann}:
\begin{equation}
{\cal M}:= {l\over 2 G} \left(e^{-2\rho} ( G^2 \pi_\rho^2 - (\eta')^2)
  +{j(\eta)\over l^2}\right)
\label{eq: define M}
\end{equation}
The surface terms $H_\pm$ are required to make the variations of the
Hamiltonian well defined.  For the given boundary conditions one finds
that \cite{BK1,GK1}
\begin{equation}
H_+ = {\sqrt{-g^+_{tt} j(\eta_+)} \over lG}\left(1- \sqrt{1-{2GMl\over
j(\eta_+)}
         } \right)
\label{eq: H+}
\end{equation}
is the quasilocal energy of the black hole in the box, whereas
$H_-= {N_0\over 2\pi} S({\cal M})$ is proportional to the thermodynamic
entropy of the black hole, expressed as a function of the mass observable.
This generalizes the recent results \cite{LW} and \cite{louko} for spherically
symmetric gravity and string inspired gravity, respectively.
\par
The Hamiltonian analysis of generic dilaton gravity has been elucidated in
numerous papers\cite{domingo2,DGK}. The only diffeomorphism invariant
observable
is the constant mode, $M$, of the mass observable, $\cal M$. Its conjugate
variable,
 $P$, is invariant
only under diffeomorphisms that vanish on the boundaries. $P$ has a
geometrical interpretation as the Killing time separation between
the two ends of the spatial slice under
consideration\cite{kuchar,DGK,domingo2,thiemann}. On the constraint
surface  the Hamiltonian
is a function only of the mass $M$ and the action of reduced theory
\cite{kuchar,BK} reads:
\begin{equation}
S[M,P] = \int dt (P\dot{M} - H(M))
\end{equation}
where again, $P$ is the time separation of the two ends of the spatial slice
and $M$ is the ADM mass of the black hole.
They obviously play the role of angle-action variables, and their naive
quantization without knowledge of the global structure of their
phase space will be as misleading as the same attempt for the harmonic
oscillator rewritten in terms of its angle-action variables. Obviously,
a correct quantization (if any) should be based on variables that
are derived from  the above angle-action ones via some canonical transformation
and give rise at the quantum level to a well defined Hilbert space with
normalizable states, etc. The recovery of these variables is certainly
not a unique procedure, and a leap of faith is needed to justify a
specific choice by relying on the universality arguments of the black
hole thermodynamics. One such procedure is as follows.
\par
Consider a canonical transformation to new variables\footnote{While this
manuscript was being prepared we were made aware of a paper by Louko and
Matela\cite{louko1} in which a similar canonical transformation was
considered.}
\begin{eqnarray}
X &=& \sqrt{B(M)/\pi} \cos ({ 2\pi P T(M)}) \\
\Pi &=& \sqrt{B(M)/\pi} \sin ({2\pi P T(M)}).
\end{eqnarray}
With arbitrary functions $B(M)$ and $T(M)$ it looks chosen ad hoc, but
gets justified in part
by noting that it is canonical if and only if the following equation
holds
\begin{equation}
\delta B= {\delta M\over T(M)} \,\,  .
\end{equation}
This equation can be interpreted as a first law of black hole thermodynamics
provided we
identify $B(M)$ with Bekenstein-Hawking entropy and $T(M)$ with the
temperature of the corresponding black hole of mass $M$:
\begin{equation}
B(M) = S(M)
\end{equation}
We have dropped the arbitrary constant and will henceforth assume
that the entropy is defined to vanish when $M=0$.
Since $X^2 + \Pi^2 = S(M)/\pi$, finding eigenstates of the mass
operator reduces to the eigenvalue problem for a simple harmonic oscillator.
Normalizability of the resulting wave function (zero boundary
conditions at the infinity of the $X$-configuration space) requires that the
eigenvalues
of $X^2 + \hat{\Pi}^2$ be $2n+1$, where $n$ is a positive integer.
This in turn yields a discrete spectrum for the entropy:
\begin{equation}
S(M) = 2\pi (n +1/2)/G
\end{equation}
as claimed.
\par
Remarkably Dirac quantization of the above system in the
Euclidean sector, for the same boundary
conditions, gives the same spectrum. As shown in \cite{BK1},
by adapting a canonical transformation first used by Cangemi {\it et al}
\cite{CJZ} for string inspired dilaton gravity, it is possible to
find exact, physical eigenstates of the mass observable in the functional
Schrodinger representation. These states take the form:
\begin{equation}
{\Psi[\rho^a,\theta_-)} = \exp\left({i\over \hbar}
{\int^{\sigma_+}_{\sigma_-}dx} \quad
  \omega(\rho^2)(\rho^0 \rho^1{}'+\rho^1\rho^0{}')\right)
  \exp({i\over \hbar}\eta_-\theta_-)
\label{eq: ansatz}
\end{equation}
where $\rho^a(x), \{a=0,1\}$ are new phase space variables (defined on
a spatial or "radial" coordinate $\sigma_-<x<\sigma_+$), conjugate to the
spatial components of the zweibein fields of the metric in the connection
representation of 2D dilaton gravity \cite{CJZ})
: $p_a = e_{a1}$. Moreoever, $\rho^2 = -(\rho^0)^2 + (\rho^1)^2 = |k|^2$.
The specific form of the functional $\omega(\rho^2)$ is not relevant for
the present discussion, except that it vanishes at the bifurcation point
given in terms of these variables by $\rho^a(\sigma_-)=0$. What is crucial,
however,
is the form of $\theta_-$, which is the variable conjugate to $\eta_-$, the
value of the dilaton at the horizon.  As shown in \cite{BK1}
in the Euclidean sector
\begin{equation}
\theta_- = \arctan (p_0/p_1)|_{\sigma_-}
\end{equation}
It is therefore periodic and singlevaluedness of the wavefunction requires
$\eta_-= n$. Given the general expression for the entropy Eq.(\ref{eq:
entropy}),
this is precisely the same quantization condition (up to an irrelevant
shift) as obtained via reduced quantization above. The positivity of $n$ does
not come from the quantum mechanics in this case. Instead it must be imposed
because $\eta=0$ corresponds to a singularity in the theory, where the
effective
gravitational constant in the theory vanishes\footnote{Alternatively, the
physical,
asymptotically flat metric has a curvature singularity.}

\section{Acknowledgements}
\par
GK grateful to E. Benedict, V. Frolov, J. Gegenberg and D. Louis-Martinez
 for helpful
discussions.  Both authors also benefitted from discussions with
J.Louko and his useful criticism. This work
was supported in part by the Natural Sciences and Engineering
Research Council of Canada. GK is grateful to the  CTP at MIT for
its hospitality during the completion of this work. AB thanks
for hospitality and financial support the
Winnipeg Institute for Theoretical Physics and the University of
Winnipeg where this work was  started. This work
was supported in part by the Natural Sciences and Engineering
Research
Council of Canada. The work of AB was also
supported by the the Russian Foundation for Basic Research under
Grants 96-02-16287 and 96-02-16295, the European
Community Grant INTAS-93-493 and the Russian Research Project
"Cosmomicrophysics".

  \par
\end{document}